\documentclass[nofootinbib,prd,aps]{revtex4}

\usepackage{graphicx}
\usepackage{dcolumn}
\usepackage{bm}

\def \be{\begin{equation}}
\def \ee{\end{equation}}

\begin{document}
\voffset = 0.3 true in

\title{Canonical Interpretation of the $D_{sJ}(2860)$ and $D_{sJ}(2690)$}

\author{F.E. Close}
\affiliation{
Rudolph Peierls Centre for Theoretical Physics, Oxford University, Oxford,
OX1 3NP, UK.}

\author{Olga Lakhina}
\affiliation{
Department of Physics and Astronomy, University of Pittsburgh,
Pittsburgh PA 15260}

\author{Eric S. Swanson}
\affiliation{
Department of Physics and Astronomy, University of Pittsburgh,
Pittsburgh PA 15260}

\author{C.E. Thomas}
\affiliation{
Rudolph Peierls Centre for Theoretical Physics, Oxford University, Oxford,
OX1 3NP, UK.}

\begin{abstract}
\vskip .3 truecm
The spectrum and decay properties of radially excited $D_s$ states are examined in a new model.
Good agreement is obtained with the properties of two recently announced $D_s$ mesons
identified as $D_{s0}(2860) = c\bar{s}(2P)$ and $D^*_{s}(2690) = c\bar{s}$ as
a possible mixture of $(2S;{}^3S_1)$ and $(1D;{}^3D_1)$.
Searching for these mesons in $B$ decays is advocated due to large predicted branching ratios.
\end{abstract}

\maketitle

\section{Introduction}  

BaBar have recently announced the discovery of a new $D_s$ state seen in $e^+e^-$ collisions 
decaying to $K^-\pi^+ K^+$, $K^-\pi^+\pi^0K^+$  ($D^0K^+$), or $D^+K^0_S$\cite{palano}. The 
Breit-Wigner mass of the new state is 

\be
M(D_{sJ}(2860)) = 2856.6 \pm 1.5 \pm 5.0 \ {\rm Mev}
\ee
and the width is
\be
\Gamma(D_{sJ}(2860)) =  48 \pm 7 \pm 10\ {\rm MeV}.
\ee
The signal has a significance greater than 5 $\sigma$ in the $D^0$ channels and 2.8 $\sigma$ in 
the $D^+$ channel. There is no evidence of the $D_{sJ}(2860)$ in the $D^*K$ decay 
mode\cite{palano} 
or the $D_s \eta$ mode\cite{bill}.

There is, furthermore, structure in the $DK$ channel near 2700 MeV that 
yields Breit-Wigner parameters of

\be
M(D_{sJ}(2690)) = 2688 \pm 4 \pm 2 \ {\rm MeV}
\ee
and
\be
\Gamma(D_{sJ}(2690)) = 112 \pm 7 \pm 36 \ {\rm MeV}.
\ee
The significance of the signal was not stated. 

The discovery of these states is particularly germane to the structure of the $D_s(2317)$. For 
example,
the low mass and isospin violating decay mode, $D_s\pi^0$, of the $D_s(2317)$ imply 
that the state could
be a $DK$ molecule\cite{BCL}. If this is the case, the $D_{sJ}(2690)$ could be a supernumerary 
scalar $c\bar s$ state. Alternatively, the $D_s(2317)$ could be the ground state scalar $c\bar s$ 
state and the
new $D_{sJ}$'s could be canonical radial excitations. Clearly, constructing a viable global model of
all
the $D_s$ states is important to developing a solid understanding of this enigmatic 
sector\cite{DsGeneral}.

Previous efforts to understand the new BaBar states have argued that the $D_{sJ}(2860)$ is a 
scalar $c \bar s$ state predicted at 2850 MeV in a coupled channel model\cite{vBR} or that it
is a $J^P = 3^-$ $c \bar s$ state\cite{CFN}.

Here we pursue a simple model that assumes that all of the known $D_s$ states are dominated by
simple 
$c\bar s$ quark content. It is known that this is difficult to achieve in the `standard' 
constituent
quark model 
with $O(\alpha_s)$ spin-dependent mass shifts
because the $D_{s0}(2317)$ is much lighter than typical predictions (for example, Godfrey
and Isgur obtain a $D_{s0}$ mass of 2480 MeV\cite{GI}). 
An essential feature in such phenomenology has been the assumption of two static potentials: a
Lorentz scalar confining potential and a short range Coulombic vector potential.  Following the 
discovery of the $D_s(2317)$, Cahn and Jackson\cite{CJ} analysed the $D_s$ states with a scalar 
potential $S$, whose shape they allowed to be arbitrary, while retaining a vector potential $V$ 
that they assumed to be Coulombic. In the limit that the mass $m_2 \gg m_1$ this enabled the spin 
dependent potential applicable to P-states to take the form

\begin{equation}
V_{SD} = \lambda L\cdot S_1 + 4\tau L\cdot S_2 + \tau S_{12}
\end{equation}
(see the discussion around Eq. 1 of \cite{CJ} for details). For $\lambda \gg \tau$ a reasonable 
description of the masses could be obtained though a consistent picture of $D_s, D$ spectroscopies 
and decays remained a problem. As the authors noted, ``the ansatz taken for the potentials $V$ and 
$S$ may not be as simple as assumed". The more general form  \cite{gromes} is

\begin{equation}
V_{SD} = \lambda L\cdot S_1 + 4\tau L\cdot S_2 + \mu S_{12}
\end{equation}
 only in the particular case of a Coulomb potential need $\mu = \tau$\cite{gromes}. Direct channel 
couplings (such as to $DK$ and $D^*K$ thresholds\cite{BCL,Rupp}) will induce effective potentials 
that allow the above more general form. Similarly, higher order gluon exchange effects in pQCD 
will also. Indeed, 
the full spin-dependent structure expected at order
$\alpha_s^2$ in QCD has been computed\cite{tye} and reveals that
an additional spin-orbit contribution to the spin-dependent interaction exists when quark masses 
are not equal. When these are incorporated 
in a constituent quark model there can be significant mass shifts leading to a lowered mass for 
the
$D_{s0}$ consistent with the $D_{s0}(2317)$\cite{LS}. Here we apply this model to the recently
discovered $D_s$ states.

\section{Canonical $c\bar s$ States}

Predictions of the new model in the $D_s$ sector are summarised in Table \ref{spectrumTab} (the
`high' parameters of  Ref. \cite{LS} are employed).

\begin{table}[!h]
\caption{$D_s$ Spectrum.}
\label{spectrumTab}
\begin{tabular}{l|cc}
\hline
\hline
state & mass (GeV)  & expt\protect\cite{PDG06} (GeV) \\
\hline
$D_s(1^1S_0)$                  &1.968 &1.968 \\
$D_s(2^1S_0)$                  &2.637 &       \\
$D_s(3^1S_0)$                  &3.097 &       \\
$D^*_s(1^3S_1)$                &2.112 &2.112 \\
$D^*_s(2^3S_1)$                &2.711 & 2.688?  \\
$D^*_s(3^3S_1)$                &3.153 &       \\
$D_s(1^3D_1)$                      &2.784 &       \\
$D_{s0}(1^3P_0)$               &2.329 &2.317 \\
$D_{s0}(2^3P_0)$               &2.817 & 2.857?\\
$D_{s0}(3^3P_0)$               &3.219 &       \\
$D_{s1}(1P)$  &2.474 &2.459 \\
$D_{s1}(2P)$  &2.940 & \\
$D_{s1}(3P)$  &3.332 & \\
$D'_{s1}(1P)$ &2.526 &2.535\\
$D'_{s1}(2P)$ &2.995 &\\
$D'_{s1}(3P)$ &3.389 &\\
$D_{s2}(1^3P_2)$               &2.577 &2.573 \\
$D_{s2}(2^3P_2)$               &3.041 & \\
$D_{s2}(3^3P_2)$               &3.431 & \\
\hline
\end{tabular}
\end{table}

Since the $D_{sJ}(2690)$ and $D_{sJ}(2860)$ decay to two pseudoscalars, their quantum numbers are 
$J^P = 0^+$, $1^-$, $2^+$, etc. Given the known states\cite{PDG06} and that the energy gap for 
radial
excitation is hundreds of MeV, on almost model independent grounds the only possibility for
a $D_{sJ}(2690)$ is an excited vector. Table \ref{spectrumTab} shows that
the $D_{sJ}(2690)$ can most naturally
be identified with the excited vector $D_{s}^*(2S)$; the D-wave vector is 
predicted to be somewhat too high at 2784 MeV though mixing between these two basis states may be 
expected.
For the $D_{sJ}(2860)$, Table \ref{spectrumTab} indicates that this is consistent with the
radially excited
 scalar state $D_{s0}(2P)$. It appears that the $D_{s2}(2P)$ is too heavy to form a viable 
identification.


\section{Decay Properties}

Mass spectra alone are insufficient to classify states. Their production and decay properties
also need to be compared with model expectations.
For example, strong decay widths can be computed with the quark model wavefunctions and the
strong decay vertex of the $^3P_0$ model. An extensive application of the model to heavy-light
mesons is presented in Ref. \cite{CS}. Here we focus on the new BaBar states with the
results given in Table \ref{StrongDecayTab}.

\begin{table}[h]
\caption{Strong Partial Widths for Candidate $D_s$ States.}
\label{StrongDecayTab}
\begin{tabular}{l|cc}
\hline
\hline
state (mass) & decay mode & partial width (MeV) \\
\hline
$D_s^*(2S)(2688)$ & $DK$   & 22 \\
            & $D^*K$ & 78 \\
            & $D_s\eta$ & 1 \\
            & $D_s^* \eta$ & 2 \\
            & total & 103 \\
\hline
$D_{s0}(2P)(2857)$ & $DK$  & 80 \\
            & $D_s\eta$ & 10 \\
            & total &  90 \\
\hline
$D_{s2}(2P)(2857)$ &  $DK$  & 3  \\
            &  $D_s\eta$ & 0 \\
            & $D^*K$ &  18 \\
            & $DK^*$ & 12 \\
            & total & 33 \\
\hline
$D_{s2}(2P)(3041)$ &  $DK$  & 1  \\
            &  $D_s\eta$ & 0 \\
            & $D^*K$ &  6 \\
            & $DK^*$ & 47 \\
            & $D^*K^*$ & 76 \\
            & total & 130 \\
\hline
\end{tabular}
\end{table}

\subsection{$D_{sJ}(2690)$}

The total width of the $D_s^*(2S)$ agrees very well with the measured width of the $D_{sJ}(2690)$ 
($112 \pm 37$ MeV), lending support
to this identification. No signal in $D_s\eta$ is seen or expected,
whereas the predicted  large $D^*K$ partial width implies that this state should be visible
in this decay mode. The data in
$D^{*0}(K) \to D^0\pi^0(K)$ do not support this contention; 
however, the modes 
$D^{*+}(K) \to D^0\gamma (K)$ and $D^{*+}(K) \to D^+\pi^0 (K)$
show indications of a broad structure near 2700 MeV\cite{palano}.
There is the possibility that $1^3D_1$ mixing with $2^3S_1$ 
shift the mass down by 30 MeV to that observed and also suppress the $D^*K$
mode. For a specific
illustration, take the model masses for the  $2^3S_1$ as 2.71GeV and $1^3D_1$ as 2.78 GeV.
A simple mixing matrix then yields a solution for the physical states with masses 2.69 GeV and its
predicted heavy partner at around 2.81 GeV with eigenstates

\begin{eqnarray}
|D_s^*(2690)\rangle &\approx& \frac{1}{\sqrt{5}}(- 2 |1S\rangle + 1 |1D\rangle) \nonumber \\
|D_s^*(2810)\rangle &\approx& \frac{1}{\sqrt{5}}(|1S\rangle + 2 |1D\rangle ) 
\end{eqnarray}
and hence a mixing angle consistent with -0.5 radians.

\begin{figure}[h]
\includegraphics[width=5 true cm, angle=270]{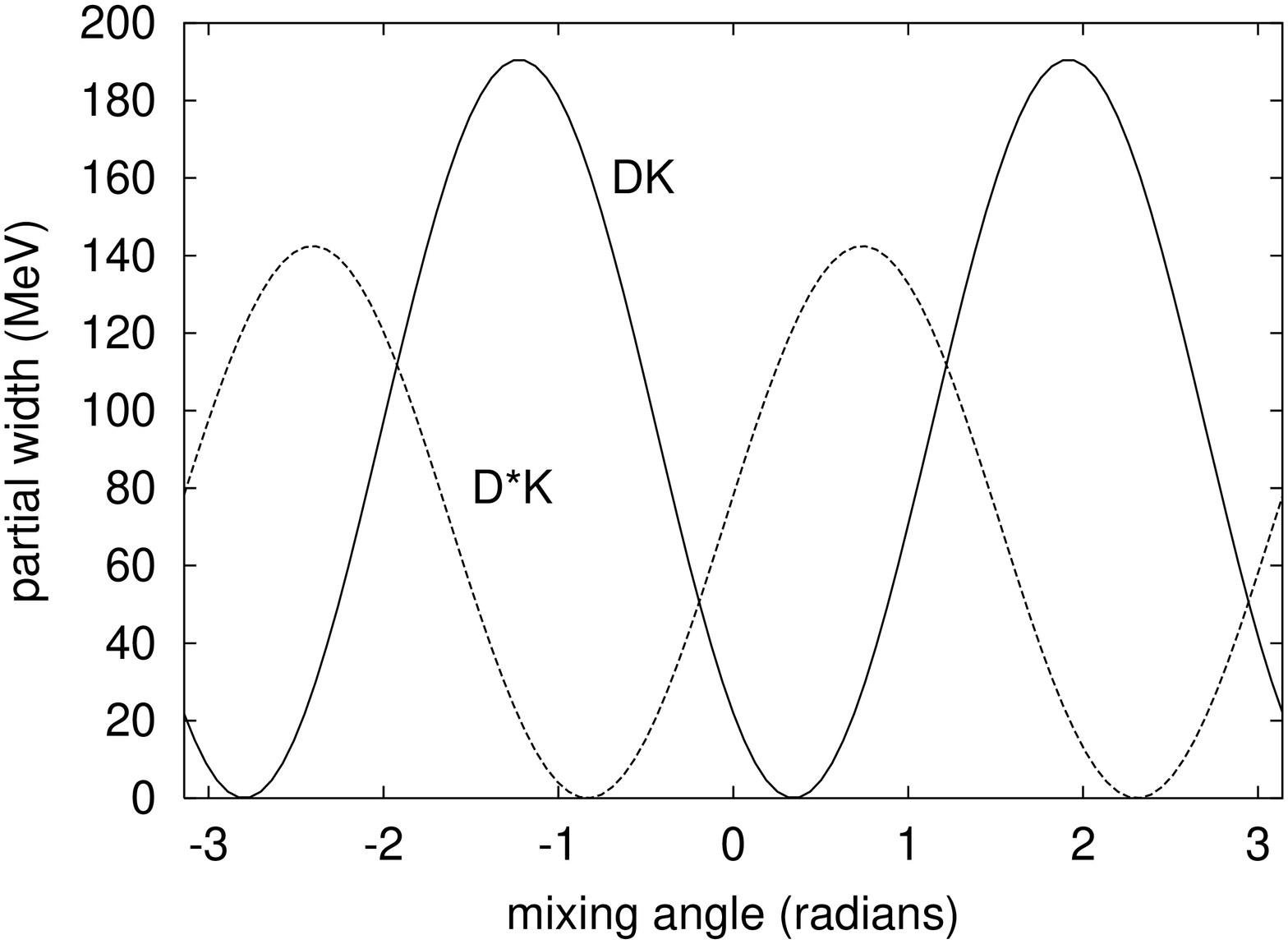}
\hskip 1 true cm
\includegraphics[width=5 true cm, angle=270]{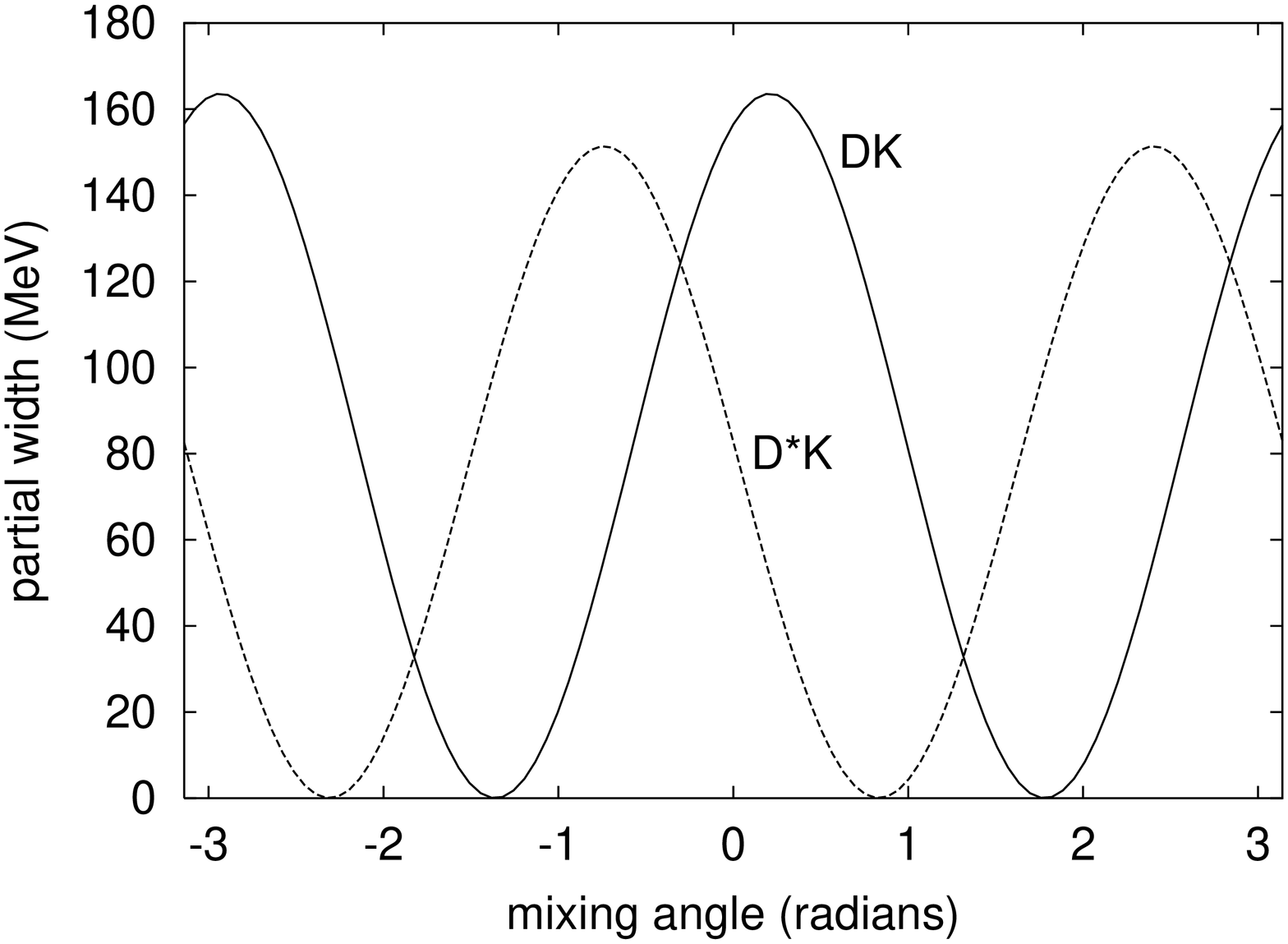}
\caption{$DK$ and $D^*K$ Partial Widths vs. Mixing Angle. Low vector (left); high vector (right).}
\label{mixedFig}
\end{figure}

The results of an explicit computation in the $^3P_0$ model are shown in  Fig. \ref{mixedFig}. 
One sees that a mixing angle of approximately -0.5 radians suppresses the $D^*K$ decay mode
of the low vector (with mass set to 2688 MeV) and produces a total width of approximately 110 MeV,
in agreement with the data. The orthogonal state would then have a mass
around 2.81 GeV and has a significant branching ratio to both $DK$ and $D^*K$, albeit with a
broad width, greater than 200 MeV.

In summary, if the $D_{sJ}(2690)$ is confirmed
as vector resonance, then signals in the $D^*K$ channel are expected, either in the low lying
state (if the mixing is weak) or in a higher vector near 2.8 GeV.

\subsection{$D_{sJ}(2860)$}

For the $D_{sJ}(2860)$, the $D_{s2}(2P)$ assignment is further disfavored. At either its model
mass of 3041 MeV or at 2860 MeV the $DK$ mode is radically suppressed, due to the
$D$-wave barrier factor. BaBar see their $D_{sJ}(2860)$ signal in $DK$ and
do not observe it in the $D^*K$  decay 
mode, making the $D_{s2}(2P)$ assignment unlikely.

By contrast, the properties of $D_{sJ}(2860)$ are consistent with
those predicted for the $D_{s0}(2P)$. Within the accuracy  
typical of the $^3P_0$ model for S-wave decays, the total width is in accord with 
the prediction that the $D_{s0}(2P)$ total width is less than that of the excited vectors,
and qualitatively in accord with the measured $48 \pm 12$ MeV.

 
\subsection{Radiative Transitions}

The meson assignments made here can be tested further by measuring radiative transitions for these
states. Predictions made with the impulse approximation, with and without nonrelativistic
reduction
of quark spinors, are presented in Table \ref{RadTransTab}.

\begin{table}[h]
\caption{$D_s$ E1 Radiative Transitions (keV).}
\label{RadTransTab}
\begin{tabular}{l|ccc}
\hline
\hline
decay mode (mass)& $q_\gamma$ (MeV) & Non Rel rate  & Rel Rate\\
\hline
$D_s^*(2S)(2688) \to D_{s0} \gamma$    & 345   & 12.7 & 4.6 \\
$D_s^*(1D)(2784) \to D_{s0} \gamma$    & 428   & 116 & 82  \\
$D_{s0}(2P)(2857) \to D_s^* \gamma$    & 648   & 13  & 0.4  \\
$D_{s2}(2P)(3041) \to D_s^* \gamma$    & 787   & 6.8  & 1.9 \\
\hline
\end{tabular}
\end{table}

\section{Production}

The production of the radially excited $D_{s0}$ in $B$ decays can be estimated with ISGW and other 
formalisms\cite{ISGW,chris}. 
Since vector and scalar $c\bar s$ states can be produced directly from the $W$ current, the decays
$B \to D_s^*(2S) D_{(J)}$ or $D_{s0}(2P) D_{(J)}$ serve as a viable source excited $D_s$ states. 
Computationally, the only
differences from ground state $D_s$ production are kinematics and the excited $D_s$ decay constants.

Production systematics can reveal structural information. For example, the decay $B^0 \to D_s^+D^-$ goes via $W$ emission with a rate proportional to $V_{bc}V_{cs}$, while $W$ exchange gives rise to $B^0 \to D_s^- K^+ \sim V_{bc}V_{ud}$ and $B^0 \to D_s^+K^- \sim V_{cd}V_{bu}$.  $W$ exchange is suppressed compared to $W$ emission, thus the expected hierarchy of rates is

\begin{equation}
\Gamma(B^0 \to D_s^+D^-) \gg \Gamma(B^0 \to D_s^- K^+) \gg \Gamma(B^0 \to D_s^+K^-).
\end{equation}
This suppression of $W$ exchange is confirmed by the data\cite{PDG06} with $BR(B^0 \to D_s^+ D^-) = (6.5\pm2.1)\times 10^{-3}$ and $BR(B^0 \to D_s^- K^+) = (3.1\pm0.8)\times 10^{-5}$.  The decay to $D_s^+ K^-$ has not been observed.

It is therefore intriguing that the observed rate for $B^0 \to D_s(2317)^+ K^-$ ($(4.3\pm1.5)\times 10^{-5}$) is comparable to $B^0 \to D_s^- K^+$.  Assuming accurate data, one must conclude either that this simple reasoning is wrong, the $D_s(2317)^- K^+$ mode will be found to be large, or the $D_s(2317)$ is an 
unusual state. Searching for the process $B^0 \to D_s(2317)^- K^+$ is clearly of great interest.

With the previous warning in mind, we proceed to analyse the production of excited $D_s$ states in a variety of models. Rates with decay constants set to 1 MeV for $D_s(2317)$ and $D_s(2860)$ production assuming that they
are simple $c\bar s$ scalar and excited scalar states are presented in Table 
\ref{table:ScalarsComparison}. 

Unfortunately, decay constants cannot be accurately computed at this time. We have evaluated 
ratios of decay constants assuming a simple harmonic oscillator quark model, a Coulomb+linear+hyperfine
quark model, and a relativised quark model. The resulting ratio for scalar mesons fall in the range
${f_{D_s(2860)} \over f_{D_s(2317)}} \approx 0.9 - 1.4$.  
The final estimates of the production of excited scalar $D_s$ mesons in $B$ decays are thus

\begin{equation}
{B \to D_s(2860) D \over B \to D_s(2317)D} = 0.6 - 1.8
\end{equation}
and
\begin{equation}
{B \to D_s(2860) D^* \over B \to D_s(2317)D^*} = 0.3 - 0.9.
\end{equation}

\begin{table}[h]
\begin{center}
\begin{tabular}{c|cccc}
\hline
{Decay Mode} & {ISGW} & {HQET - Luo \& Rosner}\protect{\cite{Rosner:HQET}} & 
{Pole}\protect{\cite{Rosner:HQET}} & {HQET - Colangelo}\protect{\cite{Colangelo:2003sa}} \\
\hline
\hline
$D_s(2317) D$ & $2.78 \times 10^{-7}$ & $1.95 \times 10^{-7}$ & $1.91 \times 10^{-7}$ & $2.24 
\times 10^{-7}$ \\
$D_s(2317) D^*$ & $1.06 \times 10^{-7}$ & $8.82 \times 10^{-8}$ & $8.79 \times 10^{-8}$ & $1.23 
\times 10^{-7}$ \\
$D_s(2860) D$ & $2.09 \times 10^{-7}$ & $1.72 \times 10^{-7}$ & $1.66 \times 10^{-7}$ & $1.83 
\times 10^{-7}$ \\
$D_s(2860) D^*$ & $4.57 \times 10^{-8}$ & $3.61 \times 10^{-8}$ & $3.55 \times 10^{-8}$ & $4.66 
\times 10^{-8}$ \\
\hline
\end{tabular}
\end{center}
\caption{Branching ratios to scalars in different models with decay constants set to 1 MeV}
\label{table:ScalarsComparison}
\end{table}


A similar analysis for vector $D_s^*$ production is presented in Table 
\ref{table:VectorsComparison}.

\begin{table}[h]
\begin{center}
\begin{tabular}{c|cccc}
\hline
{Decay Mode} & {ISGW} & {HQET - Luo \& Rosner}\protect{\cite{Rosner:HQET}} & 
{Pole}\protect{\cite{Rosner:HQET}} & {HQET - Colangelo}\protect{\cite{Colangelo:2003sa}} \\
\hline
\hline
$D_s^* D$ & $ 1.97\times 10^{-7}$ & $ 1.33\times 10^{-7}$ & $ 1.32\times 10^{-7}$ & $ 1.57\times 
10^{-7}$  \\
$D_s^* D^*$ & $ 4.20\times 10^{-7}$ & $ 3.22\times 10^{-7}$ & $ 3.23\times 10^{-7}$ & $ 4.52\times 
10^{-7}$  \\
$D_s(2690) D$ & $ 1.01\times 10^{-7}$ & $ 8.06\times 10^{-8}$ & $ 7.77\times 10^{-8}$ & $ 
8.79\times 10^{-8}$  \\
$D_s(2690) D^*$ & $ 4.66\times 10^{-7}$ & $ 3.55\times 10^{-7}$ & $ 3.49\times 10^{-7}$ & $ 
4.65\times 10^{-7}$  \\
\hline
\end{tabular}
\end{center}
\caption{Branching ratios to vectors in different models with decay constants set to 1 MeV}
\label{table:VectorsComparison}
\end{table}

Estimating vector decay constant ratios as above yields
${f_{D_s(2690)} \over f_{D_s^*}} \approx 0.7 - 1.1$.  
Finally, predicted ratios of excited vector production are

\begin{equation}
{B \to D_s(2690) D \over B \to D_s^*(2110)D} = 0.3-0.7
\label{vecRatEq}
\end{equation}
and
\begin{equation}
{B \to D_s(2690) D^* \over B \to D_s^*(2110)D^*} = 0.5-1.3.
\end{equation}
We note that Eqn. \ref{vecRatEq} agrees well with the earlier prediction of Close and Swanson\cite{CS}.

\section{Summary and Conclusions} 

Given the controversial nature of the $D_s(2317)$, establishing a consistent picture of the
entire $D_s$ spectrum is very important. The new states claimed by BaBar can be useful in this
regard. We have argued that the six known $D_s$ and two new states can be described in terms of a 
constituent quark model with novel spin-dependent interactions. Predicted strong decay properties 
of these states appear to agree with experiment. 

Perhaps the most important tasks at present are (i) discovering the $D_{s2}(2P)$ state, (ii) 
searching for resonances in $D^*K$ and $DK^*$ up to 3100 MeV, (iii) analysing the angular 
dependence
of the $DK$ final state in $D_{sJ}(2860)$ decay, (iv) assessing whether the $D_{sJ}(2690)$ 
appears in the $D^*K$ channel, (v) searching for these states in $B \to D_{sJ}D^{(*)}$ with
branching ratios of $\sim 10^{-3}$.


\subsection{Postscript: Belle discovery}

Subsequent to these calculations, and as this report was being completed, Belle\cite{newBelle} 
has reported a vector state whose mass, width, and possibly production rate and decay
characteristics are consistent with our
predictions. Specifically, their measured mass and total width are
$M = 2715 \pm 11 ^{+11}_{-14}$ MeV and $\Gamma = 115 \pm 20 ^{+36}_{-32}$
MeV, in remarkable agreement with our predictions. The specific parameters we have used in 
our analysis are contained within their uncertainties.

Belle\cite{newBelle} find the new state in $B$ decays, which we have proposed as a likely source. 
They report
$Br(B \to \bar{D^0}D^*_{s}(2700))\times Br(D^*_{s}(2700) \to D^0K^+) =
(7.2 \pm 1.2 ^{+1.0}_{-2.9})\cdot 10^{-4}$. When compared to the production of the ground state
vector\cite{PDG06} which is $Br(B \to \bar{D^0}D^*_{s}(2112)) = (7.2 \pm 2.6)\cdot 10^{-3}$, the 
ratio of production rates in $B$ decay is then ${\cal{O}}(0.1)/Br(D^*_{s}(2700) \to D^0K^+)$. 
From our Table II, and assuming flavor symmetry for the strong decay, we predict that
$Br(D^*_{s}(2690) \to D^0K^+) \sim 10\%$ , which within the uncertainties will apply also to
the Belle state. Thus the absolute production rate, within the large uncertainties, appears to be 
consistent with that predicted in Section 4. If the central value of the Belle mass is a true
guide, then a significant branching ratio in $D^*K$ would be expected (Table II and Fig 1).
The orthogonal vector state would then be dominantly 1D at 2.78 GeV, but hard to produce in
$B$ decays.  These statements depend on the dynamics underlying $2S$-$1D$ mixing, which is poorly
understood. It is therefore very useful that $B$ decay systematics and the strength of the $D^*K$
decay channel in the excited vector $D_s$ mesons can probe this dynamics.


Searching for this state in the other advocated modes, and improving the uncertainties, now
offers a significant test of the dynamics discussed here.

\acknowledgments
This work is supported, in part, by grants from the Particle Physics
and Astronomy Research Council, the EU-TMR program ``Euridice''
HPRN-CT-2002-00311 (Close and Thomas),
and the U.S. Department of Energy under contract DE-FG02-00ER41135 (Swanson and Lahkina).

\end{document}